\documentclass[10pt,twocolumn,letterpaper]{article}

\usepackage{cvpr}              % To produce the CAMERA-READY version
\usepackage{multirow}
\usepackage{array}
\usepackage{booktabs} % 三线表
\usepackage{multirow} % 多行合并
\usepackage{graphicx} % 缩放
\usepackage[table]{xcolor} % 表格颜色支持
\usepackage{bigstrut} % 调整行高
\usepackage{tabularx}
\definecolor{tableBlue}{rgb}{0.424,0.557,0.749}
\definecolor{lightblue}{HTML}{dfebf7}
\newcommand{\tableLineBlue}{\rowcolor{lightblue}}

\usepackage{arydshln}
\definecolor{cvprblue}{rgb}{0.21,0.49,0.74}
\usepackage[pagebackref,breaklinks,colorlinks,allcolors=cvprblue]{hyperref}

\def\paperID{16960} % *** Enter the Paper ID here
\def\confName{CVPR}
\def\confYear{2026}
\vspace{-2ex}
\title{ProAV-DiT: A Projected Latent Diffusion Transformer for Efficient Synchronized Audio-Video Generation}
\vspace{-2ex}
\author{
Jiahui Sun\textsuperscript{1} \quad
Weining Wang\textsuperscript{1} \quad
Mingzhen Sun\textsuperscript{1} \quad
Yirong Yang\textsuperscript{2} \\
Xinxin Zhu\textsuperscript{1} \quad
Jing Liu\textsuperscript{1} \\
\\
\textsuperscript{1}Institute of Automation, Chinese Academy of Sciences \\
\textsuperscript{2}Beihang University \\
\\
}
\begin{document}
\maketitle
\vspace{-2ex}
\begin{abstract}
%Sounding video generation (SVG) has emerged as a challenging task due to the inherent cross-modal temporal and semantic misalignment and the high computational costs associated with multimodal data. To address these issues, we propose the Projected Latent Audio-Video Diffusion Transformer (ProAV-DiT), a novel diffusion transformer explicitly designed for synchronized audio-video synthesis. Our approach introduces a Multi-scale Dual-stream Spatio-temporal Autoencoder (MDSA) that bridges audio and video modalities through a unified cross-modal latent space.  This framework compresses audio and video inputs into 2D latents, each capturing distinct aspects of the signals. To further enhance audiovisual consistency and facilitate cross-modal interaction, MDSA incorporates a multi-scale attention mechanism that enables temporal alignment across resolutions and supports fine-grained fusion between modalities. To effectively capture the fine-grained spatiotemporal dependencies inherent in SVG tasks, we introduce the Spatio-Temporal Diffusion Transformer (STDiT) as the generator of our framework. Extensive experiments demonstrate that our method achieves superior performance on standard benchmarks (Landscape and AIST++), outperforming existing methods on all evaluation metrics on the Landscape dataset while significantly improving training and sampling speed. We further explore its performance on open-domain SVG on AudioSet, proving the generalization ability of ProAV-DiT.
Sounding Video Generation (SVG) remains a challenging task due to the inherent structural misalignment between audio and video, as well as the high computational cost of multimodal data processing. In this paper, we introduce ProAV-DiT, a Projected Latent Diffusion Transformer designed for efficient and synchronized audio-video generation. To address structural inconsistencies, we preprocess raw audio into video-like representations, aligning both the temporal and spatial dimensions between audio and video. At its core, ProAV-DiT adopts a Multi-scale Dual-stream Spatio-Temporal Autoencoder (MDSA), which projects both modalities into a unified latent space using orthogonal decomposition, enabling fine-grained spatiotemporal modeling and semantic alignment. To further enhance temporal coherence and modality-specific fusion, we introduce a multi-scale attention mechanism, which consists of multi-scale temporal self-attention and group cross-modal attention. Furthermore, we stack the 2D latents from MDSA into a unified 3D latent space, which is processed by a spatio-temporal diffusion Transformer. This design efficiently models spatiotemporal dependencies, enabling the generation of high-fidelity synchronized audio-video content while reducing computational overhead. Extensive experiments conducted on standard benchmarks demonstrate that ProAV-DiT outperforms existing methods in both generation quality and computational efficiency.
\end{abstract}
    
\section{Introduction}
\label{sec:intro}
Sounding Video Generation (SVG) is a multimodal content generation task that aims to synthesize dynamic videos directly from static images or textual inputs, while simultaneously generating semantically aligned and temporally synchronized audio. 
Its ability to produce coherent audio-video content makes SVG promising for applications in film production, virtual reality, and intelligent media.
Existing SVG methods \citep{liu2023sounding, yang2025cmmd, yariv2024diverse, AV-DiT, zhao2025uniform, ruan2023mm, sun2024mm, xing2024seeing} can be broadly categorized into two groups. The first is cascade generation, where video is generated first, followed by audio conditioned on the video. 
These methods often rely on additional alignment modules to mitigate synchronization issues \citep{xing2024seeing, zhang2024foleycrafter, comunita2024syncfusion, tpos, yu2024languagemodelbeatsdiffusion}, but typically suffer from temporal misalignment and error accumulation across cascaded stages. 
The second is synchronized generation, where both audio and video are generated jointly within a unified framework. 
While this approach reduces global alignment errors, achieving fine-grained synchronization and semantic coherence remains challenging. Representative methods \citep{ruan2023mm, AV-DiT, sun2024mm, liu2023sounding} are either trained directly in the signal space, incurring high computational costs, or adopt image DiT, which lack the capacity for fine-grained spatiotemporal modeling.
Despite these efforts, fine-grained spatiotemporal alignment remains an open challenge, primarily due to the intrinsic heterogeneity between audio and video.

\begin{figure}[t]
    \centering
    \includegraphics[width=1.0\linewidth]{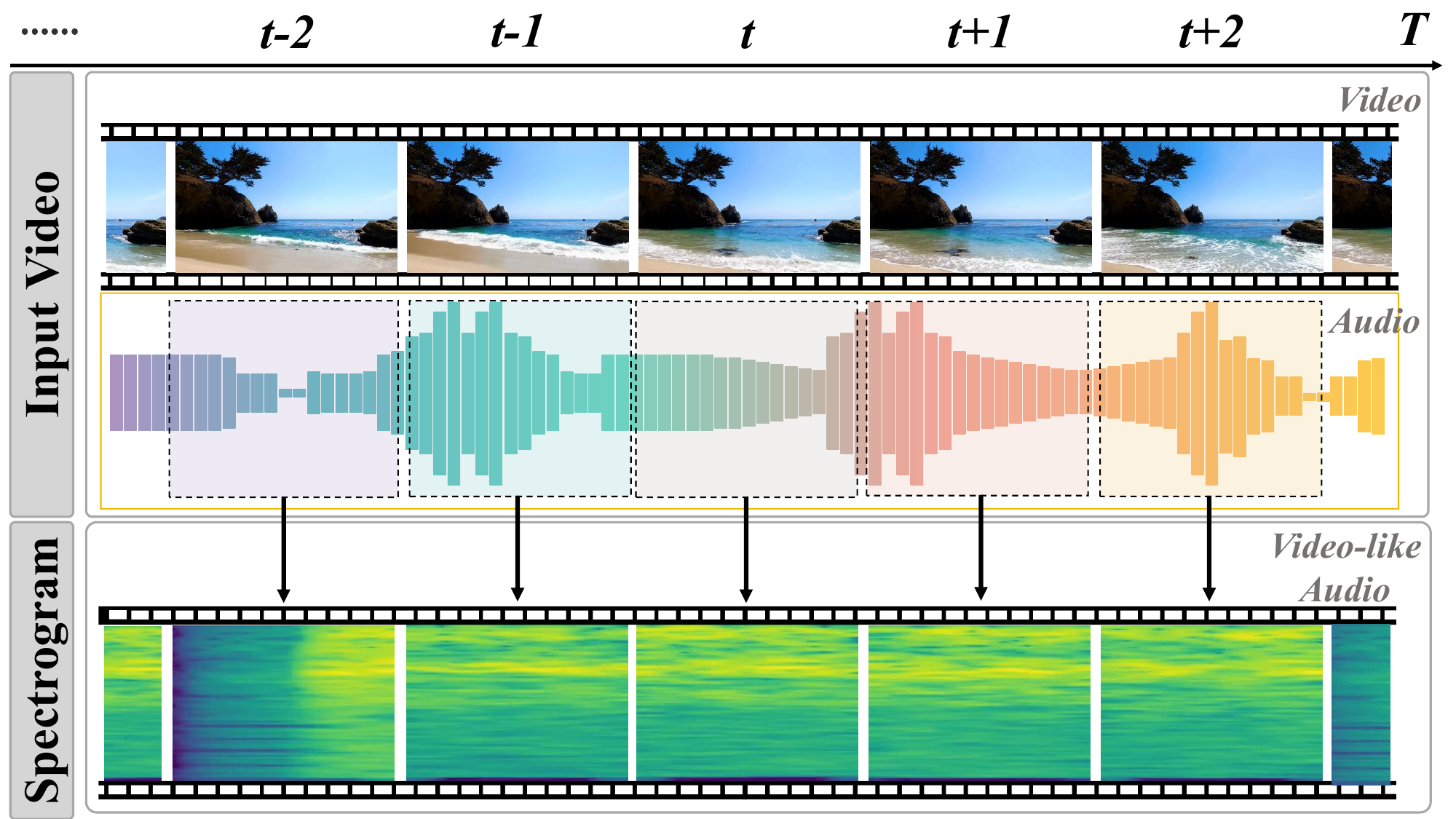}
    \caption{Video-like audio representation construction. The audio is segmented by frame, divided into audio segments in each colored square, and then converted into the Mel spectrogram sequence below. Each spectrogram has the same duration as the video frame, and the sequence is stacked along the time dimension to form a video-like audio representation ($A \in \mathbb{R}^{T \times H \times W}$), where the spectrogram acts as an image-like frame.}
    \label{video-like audio}
    \vspace{-3ex}
\end{figure}

To systematically characterize the limitations of the SVG task, we identify three fundamental challenges that distinguish SVG from conventional unimodal video generation.
\textbf{First}, the structural heterogeneity of audio and visual data introduces significant modeling difficulties. 
Video is a 3D tensor with dimensions for time, height, and width, necessitating complex spatiotemporal modeling. In contrast, audio is a 1D waveform with much higher temporal resolution (22.7 $\mu s$ vs. 33.3 $ms$ per video frame), resulting in a 1500$\times$ mismatch. This disparity complicates latent space alignment and hinders joint optimization. 
\textbf{Second}, achieving temporal consistency between audio and visual modalities remains highly challenging. For instance, synchronizing an explosion’s sound with its corresponding visual impact requires precise multi-scale temporal modeling. However, existing methods \citep{ruan2023mm, AV-DiT} often rely on shallow or global strategies, thus struggling to capture the fine-grained temporal dependencies essential for accurate synchronization.
\textbf{Third}, computational inefficiency remains a critical obstacle for large-scale synchronized audio-video generation. High-dimensional inputs and multi-branch architectures incur substantial memory and computation costs, severely limiting scalability and practical deployment.

To address these challenges, we propose the Projected Latent Audio-Video Diffusion Transformer (ProAV-DiT), a unified framework for efficient SVG. 
\textbf{First}, to mitigate the structural inconsistency between audio and video, we preprocess raw audio into video-like audio representations. As depicted in Fig.~\ref{video-like audio}, audio is segmented into frame-wise chunks, converted to mel-spectrograms, and stacked along the temporal axis.
This process aligns the structure of audio with that of video, enabling them to share a compatible format and facilitating unified encoding. 
\textbf{Second}, we introduce the Multi-scale Dual-stream Spatio-Temporal Autoencoder (MDSA) for fine-grained spatiotemporal modeling and semantic alignment. 
MDSA employs orthogonal decomposition to compress both audio and video into 2D latent representations along three axes (temporal, height, and width), disentangling dynamic and static content while reducing dimensionality.
It further incorporates a multi-scale attention mechanism, improving temporal coherence through Multi-scale Temporal Self-Attention (MT-SelfAttn) and aligning modality-specific features using Group Cross-Modal Attention (GCM-Attn). Finally, Bidirectional Block Cross-Attention (Bi-Block CrossAttn) enhances semantic alignment across modalities in the decoder.
\textbf{Third}, to improve computational efficiency, we stack the 2D latents from MDSA into a unified 3D latent and feed it into STDiT, which efficiently models spatiotemporal dependencies via serialized spatial–temporal attention. Operating in latent space greatly reduces memory and computation, enabling high-fidelity audio–video synthesis.

We validated the effectiveness and efficiency of our method through extensive experiments on the Landscape \citep{lee2022sound} and AIST++ \citep{li2021ai} datasets, and the open-domain AudioSet \citep{7952261}. Empirical results show our model achieves competitive performance and significantly improved efficiency against established benchmarks. Notably, on the Landscape dataset, our model substantially improves the FVD score by $23.5\%$ while accelerating sampling speed by $2.2\times$ over the leading comparable method.
Our contributions are summarized as follows:
\begin{itemize}
    \item We propose ProAV-DiT, a unified framework for synchronized audio-video generation, incorporating video-like audio representations to align audio and video in a shared latent space, improving cross-modal alignment and temporal coherence.
    \item We design MDSA, which introduces orthogonal feature decomposition to disentangle spatial and temporal components, reducing redundancy and enabling efficient yet expressive cross-modal fusion.
    \item We design a multi-scale attention mechanism, consisting of MT-SelfAttn for temporal modeling, GCM-Attn for modality-specific fusion, and Bi-Block CrossAttn for localized cross-modal integration, enhancing motion consistency and synchronization.
    \item Extensive evaluations demonstrate that ProAV-DiT achieves superior generation quality and inference efficiency. Its effectiveness on a large-scale open-domain dataset further validates strong generalization capabilities.
\end{itemize}

\vspace{-1ex}
\section{Related Work}
\label{sec:related}
\subsection{Diffusion Models}
Diffusion Models (DMs) \citep{ddpm, rombach2022high} have shown strong performance in image, video, and audio generation through iterative denoising. Extending DMs to video generation \citep{singer2022make, ho2022video, zhou2022magicvideo, he2022lvdm, videofactory, ho2022imagenvideohighdefinition} raises challenges in spatiotemporal modeling and computational cost.
VDM \citep{ho2022video} utilizes 3D convolutions but suffers from high computational overhead. Make-A-Video \citep{singer2022make} and VideoLDM \citep{he2022lvdm} decouple spatial-temporal modeling via 2D/1D convolutions and 3D-VAEs. Imagen Video \citep{ho2022imagenvideohighdefinition} and PVDM \citep{pvdm} reduce computational costs by applying latent compression, with PVDM using 2D latents. However, these methods are still mainly focused on unimodal video generation.
Most diffusion models are based on the Transformer. Diffusion Transformers (DiTs) \citep{dit, sora} replace U-Net with global self-attention to better capture long-range dependencies. Latte \citep{ma2024latte}, CogVideoX \citep{yang2024cogvideox}, Sora \citep{sora}, and VDT \citep{walt} improve scalability and temporal alignment through latent 3D blocks. Despite these advances, most DiTs remain modality-specific, lacking effective integration between audio and video.
In contrast, our framework unifies audio and video through aligned representations and reduces complexity via hierarchical feature decoupling.

\subsection{Sounding Video Generation (SVG)}
Unlike silent video generation, SVG requires the synchronous synthesis of high-quality audio-video content. Existing methods can be broadly classified into two categories.The first is cascade generation \citep{zhang2024letstalk, zhang2024foleycrafter, xing2024seeing,yang2025cmmd}, while the second is synchronized generation \citep{liu2023sounding, yariv2024diverse, AV-DiT, ruan2023mm, sun2024mm, zhao2025uniform}.
MM-Diffusion \citep{ruan2023mm} is a pioneering method that employs diffusion models to jointly generate audio and video. 
It introduces two pairs of denoising diffusion models for synchronized generation and proposes a randomly shifted attention mechanism to model cross-modal consistency.
MM-LDM \citep{sun2024mm} is the first latent diffusion model specifically designed for SVG, mapping audio and video to a shared semantic space via a hierarchical multimodal autoencoder. 
AV-DiT \citep{AV-DiT} employs a shared pre-trained DiT backbone with lightweight adapter modules, allowing for the adaptation of a frozen image generator to audio-video tasks while reducing computational overhead.
Uniform \citep{zhao2025uniform} employs DiT to integrate visual and audio tokens into a unified latent space, enabling joint representation learning and audio-video generation.
Despite recent advances, existing SVG methods still face challenges such as coarse temporal alignment, limited modality correspondence, or rigid architecture designs. Some approaches rely on parameter sharing or global latent alignment \citep{AV-DiT, sun2024mm}, while others process raw audio or spectrograms without preserving fine-grained audiovisual consistency \citep{liu2023sounding, ruan2023mm, lee2023aadiff}. 
In contrast to prior methods, we propose ProAV-DiT with hierarchical alignment for improved multi-scale temporal modeling. We further enhance cross-modal fusion by aligning modality-homogeneous latent spaces at multiple scales.

\begin{figure*}[h]
    \centering
    % 下方的图像
    \begin{subfigure}[t]{\textwidth}
        \centering
        \includegraphics[width=0.95\textwidth]{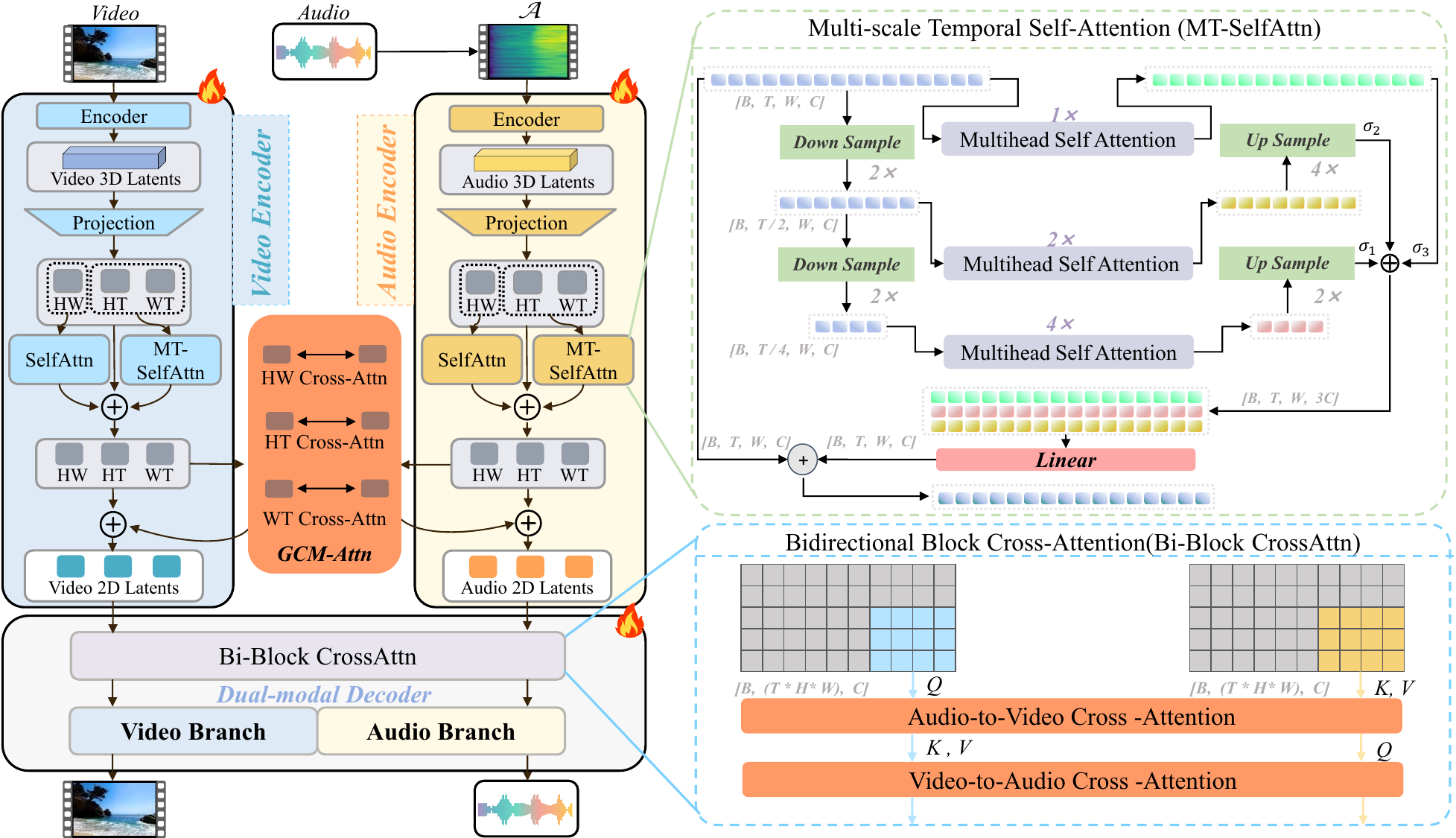}
        \caption{Overview of MDSA architecture.. }
        \label{mdsa}
    \end{subfigure}
    % 添加一些垂直间距
    \vspace{3pt}
    % 上方的图像
    \begin{subfigure}[t]{\textwidth}
        \centering
        \includegraphics[width=0.95\textwidth]{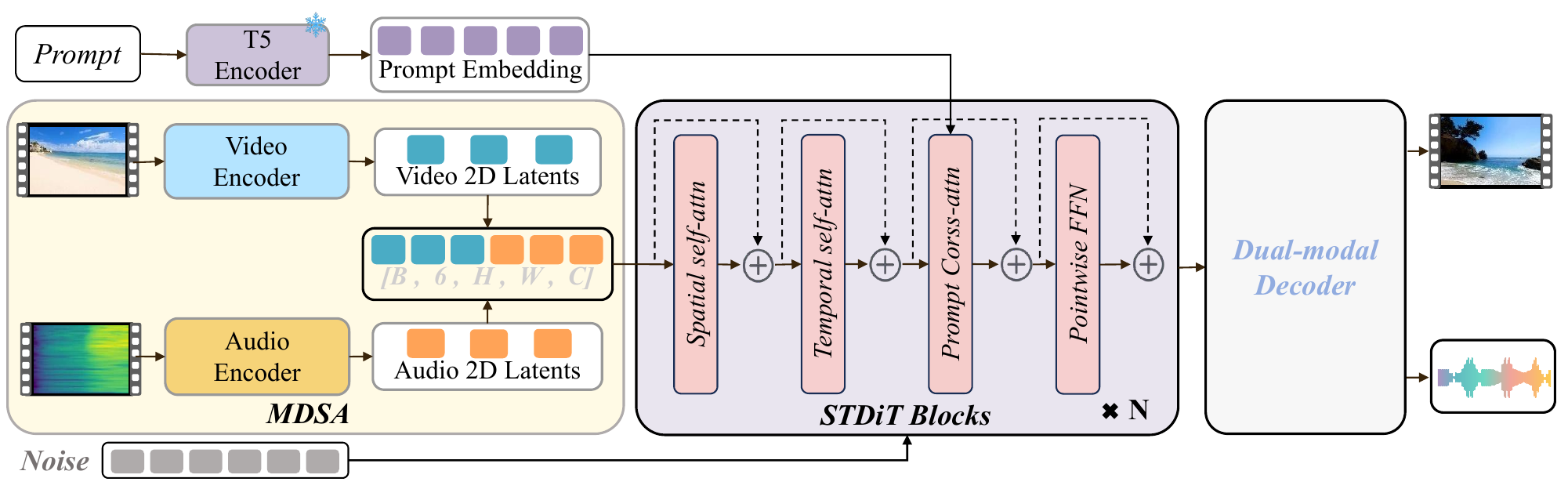}
        \caption{The detailed architecture of ProAV-DiT.}
        \label{ProAV-DiT}
    \end{subfigure}
    % 可选的总体标题
    \vspace{-2ex}
    \caption{(a) Given audio and video inputs, the audio is first converted into a video-like representation ($\mathcal{A}$). Both modalities are encoded via video-to-3D-latent encoders and projected into 2D latents through orthogonal decomposition. These latents are enhanced and fused using a multi-scale attention mechanism: temporal consistency (HT, WT) is modeled by MT-SelfAttn, spatial features (HW) are refined by SelfAttn, and GCM-Attn enables bidirectional cross-modal interaction. The resulting 2D latents are further processed by Bi-Block CrossAttn and decoded by a dual-modal decoder to produce synchronized audio-video outputs.
    (b) Audio and video latents are concatenated along the temporal axis to form a unified 3D latent representation, which serves as input to the ST-DiT. During iterative diffusion, ST-DiT progressively denoises the latents at each timestep. After the final step, the purified latents are decoded to synthesize video with temporally aligned audio-video streams.
    }
    \vspace{-2ex}
    \label{fig:architectures}
\end{figure*}

\section{Method}
This section introduces ProAV-DiT, a novel framework for synchronized audio-video generation.
To tackle the challenges of cross-modal alignment and computational efficiency, ProAV-DiT integrates two components: a Multi-scale Dual-stream Spatio-temporal Autoencoder (MDSA) for encoding audio and video into a unified latent space, and an Audio-Video Diffusion Transformer for generating synchronized content in that space. The overall framework is illustrated in Fig.~\ref{ProAV-DiT}.

\subsection{Multi-scale Dual-stream Spatio-temporal Autoencoder (MDSA)} 
\label{sec:mdsa}
The MDSA processes video and audio inputs to create aligned, low-dimensional latent representations. As shown in Fig.~\ref{mdsa}, the MSDA processes both the video tensor $\mathcal{V} \in \mathbb{R}^{T \times H \times W}$ and the audio tensor $\mathcal{A} \in \mathbb{R}^{ T \times H \times W}$ through three collaborative stages: dual-stream encoder, multi-scale attention mechanism, and dual-modal decoder.
The encoder decomposes each modality into compact $2D$ latent representations, suitable for the subsequent diffusion process. The multi-scale attention mechanism enhances temporal coherence and facilitates cross-modal fusion, thereby improving semantic alignment and synchronization. The decoder reconstructs synchronized video and audio from the fused latent representations.
\vspace{-1ex}

\paragraph{Input Preprocessing.}
To align the structural differences between modalities, raw audio is segmented into frame-wise chunks, converted into mel-spectrograms, and stacked along the temporal axis to match the video’s temporal structure.
This results in a video-like audio representation $\mathcal{A} \in \mathbb{R}^{T \times H \times W}$,  where each spectrogram acts as an image-like frame synchronized with the corresponding video frame, as illustrated in Fig.~\ref{video-like audio}.
Unlike MM-LDM, our method supports direct reconstruction via inverse Mel transformation, avoiding reliance on neural vocoders like HiFi-GAN \citep{hifigan} and reducing conversion errors.
\vspace{-2ex}

\paragraph{Dual-stream Encoder.}
The dual-stream encoder plays a crucial role in processing the audio and video inputs separately before they are fused. Unlike traditional methods that process video data as a 3D tensor, our approach decomposes each modality into compact 2D latent representations through orthogonal decomposition, which enhances efficiency and maintains temporal coherence by explicitly disentangling spatiotemporal factors. This decomposition is pivotal for cross-modal alignment, as it reduces redundancy and isolates axis-specific features, thereby simplifying the alignment process between audio and video modalities.
Specifically, given an audio or video representation $x$, we compute a set of disentangled 2D latents $\mathbf{z} = [\mathbf{z}^t, \mathbf{z}^h, \mathbf{z}^w]$ using an encoder $f_\phi$,
which consists of a video-to-3D encoder and three 3D-to-2D projectors. The encoding process can be formulated as:
\begin{equation}
    \text{u} :=  f^{\text{thw}}_{\phi_{\text{thw}}}(x), \text{where u} \in \mathbb{R}^{T \times H' \times W'}
\end{equation}
\begin{equation}
    \mathbf{z} = [\mathbf{z}^t, \mathbf{z}^h, \mathbf{z}^w] \quad \text{where} \quad
    \begin{cases}
    \mathbf{z}^t = \mathbf{z}_{hw} = f^t_{\phi_t}(u) \in \mathbb{R}^{H' \times W'} \\ 
    \mathbf{z}^h = \mathbf{z}_{tw} = f^h_{\phi_h}(u) \in \mathbb{R}^{T \times W'} \\ 
    \mathbf{z}^w = \mathbf{z}_{wh} = f^w_{\phi_w}(u) \in \mathbb{R}^{T \times H'}\\
    \end{cases}
\end{equation}
where $f^{\text{thw}}_{\phi_{\text{thw}}}$ is a video-to-3D-latent encoder, $f_\phi^t$, $f_\phi^h$, and  $f_\phi^w$ are the 3D to 2D projection modules, 
% where 
$H' = H/d$ and $W' = W/d$ denote the downsampled spatial dimensions, and $T$ represents the number of temporal segments. Specifically, $\mathbf{z}_t$ encodes shared temporal information between video and audio, such as video backgrounds and audio spectral features, while $\mathbf{z}_h$ and $\mathbf{z}_w$ capture motion patterns along the height and width axes of the video, respectively. Our design is inspired by tensor decomposition and multi-view learning, where high-dimensional spatiotemporal data can often be approximated as sums of separable components:
\begin{equation}
\mathbf{u} \approx \sum_{r=1}^{R} \mathbf{a}_r \otimes \mathbf{b}_r \otimes \mathbf{c}_r,
\end{equation}
with $\otimes$ denoting the outer product, and $\mathbf{a}_r \in \mathbb{R}^T, \mathbf{b}_r \in \mathbb{R}^{H'}, \mathbf{c}_r \in \mathbb{R}^{W'}$ representing variation along temporal, height, and width axes. Instead of performing full factorization, we extract the three structured projections $(\mathbf{z}^t, \mathbf{z}^h, \mathbf{z}^w)$, yielding complementary, low-redundancy latent representations. 
%%%%rebuttal
The orthogonal decomposition aids alignment by minimizing cross-axis redundancy, which allows for independent processing of each axis. This axis-wise disentanglement enables the model to align audio and video features along corresponding axes without interference from irrelevant dimensions. We measure the degree of independence between axes by calculating the pairwise mutual information (MI) between the latent representations. Lower MI values indicate a higher degree of decoupling (detailed in Appendix~\ref{MI}). From an information theory perspective, this decomposition not only minimizes redundancy but also maximizes the mutual information between corresponding audio-video latents along each axis, providing a theoretical basis for efficient, fine-grained cross-modal alignment.

\paragraph{Multi-scale attention mechanism.}
To ensure cross-modal consistency between audio and video, we perform temporal modeling and feature interaction on the spatiotemporal orthogonal representations $\mathbf{z}_{tw} \in \mathbb{R}^{T \times W}$ and $\mathbf{z}_{th} \in \mathbb{R}^{T \times H}$.
For decoupled spatiotemporal modeling, spatial self-attention is applied to the static content $\mathbf{z}_{hw}$, enhancing spatial feature representations and capturing long-range dependencies across different locations within each video frame.
The architecture employs spatial self-attention for spatial components and introduces MT-selfAttn for temporal modeling. 
MT-selfAttn follows a two-step process. \textbf{First}, 2$\times$ and 4$\times$ average pooling is applied along the temporal axis $T$ to extract multi-scale features at three different temporal resolutions.
\begin{equation}
    \begin{aligned}
        \mathbf{z}^{(1)}  &= \mathbf{z}_t \in \mathbb{R}^{T \times S}\\
        \mathbf{z}^{(2)}  &= \text{AvgPool}_{2\times}(\mathbf{z}_t) \in \mathbb{R}^{T/2 \times S} \\ 
        \mathbf{z}^{(3)}  &= \text{AvgPool}_{4\times}(\mathbf{z}_t) \in \mathbb{R}^{T/4 \times S} \\ 
    \end{aligned}
\end{equation}
where $S$ represents the spatial dimension ($H'$ or $W'$), and $\mathbf{z}_t$ corresponds to either $\mathbf{z}_{th}$ or $\mathbf{z}_{tw}$.
Self-attention is then applied to the features at each scale, with global-scale attention preserving temporal consistency, mid-scale attention capturing semantic relationships, and local-scale attention detecting transient patterns.
\textbf{Second}, the attended features are upsampled back to the original temporal resolution and aggregated as follows:
\begin{equation}
\tilde{\mathbf{z}}_t = \sum_{n \in \{\text{1,2,3}\}}\text{DeConv}_{k_n}\left(\text{Attn}_t(\mathbf{z}^{(n)})\right)
\end{equation}
where $\text{DeConv}_{k_n}$ uses upsampling rate $k_n \in \{1,2,4\}$ to restore the original resolution.

The architecture hierarchically encodes visual and audio features, using decoupled spatial and temporal self-attention to capture intra-modal dependencies efficiently. 

To facilitate fine-grained cross-modal fusion, we employ GCM-Attn between corresponding 2D latents from the audio and video branches. Let $\mathbf{z}_v$ and $\mathbf{z}_a$ denote video and audio latent features, respectively, derived from the three orthogonal axes $\mathbf{z}^t$, $\mathbf{z}^h$, and $\mathbf{z}^w$. For each group $\mathbf{z}$, the audio latent is updated as follows:
\begin{equation}
\mathbf{z}_a = \text{CA}(\mathbf{z}_v,\mathbf{z}_a) + \mathbf{z}_a
  \text{,}\ \mathbf{z} \in [\mathbf{z}^t, \mathbf{z}^h, \mathbf{z}^w]
\end{equation}
Here, $\text{CA}(\cdot)$ represents a cross-attention module where the query originates from the audio latent $\mathbf{z}_a$ and the key-value pair is sourced from the corresponding video latent $\mathbf{z}_v$. The residual connection preserves modality-specific content while integrating semantically aligned video context. In a symmetric manner, the video latents are updated using the audio latents as keys and values, thereby enabling bidirectional information exchange. 
By combining multi-scale temporal modeling with independent spatial and temporal processing, our framework mitigates feature entanglement, eliminates redundant computation, and ensures efficient, semantically coherent cross-modal alignment.
\vspace{-1ex}

\paragraph{Dual-modal Decoding Architecture.}
During decoding, the dual-modal decoder reconstructs both video and audio streams using a dual-branch architecture (Fig.~\ref{mdsa}), enabling unified multimodal modeling. The features $\mathbf{z}_{a}'$ and $\mathbf{z}_{v}'$ are obtained by first applying the encoder for feature disentanglement, followed by a multi-scale attention mechanism to enhance the temporal coherence and semantic expressiveness of the latent representations.
\begin{equation}
\mathbf{Z}_{t \times h \times w} = \mathcal{E}_{t}(\mathbf{z}_{hw}') + \mathcal{E}_{h}(\mathbf{z}_{tw}') + \mathcal{E}_{w}(\mathbf{z}_{th}') \in \mathbf{R}^{T \times H \times W}
\end{equation}
where the $\mathcal{E}$ operator expands feature maps along temporal, height, and width dimensions for alignment and fusion. 
Bi-Block CrossAttn is applied to model cross-modal interactions across spatiotemporal dimensions. To ensure efficiency and semantic relevance, the unified latent tensor is divided into non-overlapping blocks. Within each block, cross-attention is computed between audio and video features to capture localized correlations. The attention matrix $\mathbf{A}$ is defined as:
\begin{equation}
\mathbf{A}_{i,j} = \frac{\exp(\mathbf{q}_i^\top \mathbf{k}_j / \sqrt{d})}{\sum_{k} \exp(\mathbf{q}_i^\top \mathbf{k}_k / \sqrt{d})},
\end{equation}
where $\mathbf{q}_i$ and $\mathbf{k}_j$ are query and key vectors from the video and audio modalities, respectively. $\mathbf{A}_{i,j}$ indicates how the $i$-th video patch attends to the $j$-th audio patch, enabling block-wise semantic alignment. This design supports adaptive context modeling while maintaining efficiency through sparse attention.

\subsection{Cross-Modal Diffusion Transformer for Synchronized Generation}
As illustrated in Fig.~\ref{ProAV-DiT}, ProAV-DiT first encodes video and audio inputs into three orthogonal 2D latent representations using the MDSA encoder: $\mathbf{z}_v = [\mathbf{z}^t_v, \mathbf{z}^h_v, \mathbf{z}^w_v]$ for video and $\mathbf{z}_a = [\mathbf{z}^t_a, \mathbf{z}^h_a, \mathbf{z}^w_a]$ for audio. These representations respectively capture spatiotemporal visual features and time-frequency audio characteristics. 
To enhance temporal consistency in modeling the joint data distribution $p_{\text{data}}(\mathbf{z}_v, \mathbf{z}_a)$, we adopt a temporal stacking strategy within the STDiT framework, where modality-specific latent tensors are concatenated along the temporal dimension to form a unified six-frame latent sequence. 
\begin{equation}
    \mathbf{P} = \text{Stack}([\mathbf{z}^t_v, \mathbf{z}^h_v, \mathbf{z}^w_v, \mathbf{z}^t_a, \mathbf{z}^h_a, \mathbf{z}^w_a]) \in \mathbb{R}^{6 \times H \times W}
\end{equation}
The $\mathbf{P}$ serves as input to the STDiT generator, which jointly models cross-modal interactions. The resulting latent outputs are then decoded by a dual-modal decoder to synthesize videos with temporally aligned audio and video content.
The framework enables efficient multimodal fusion through two key innovations. First, we represent the audio and video latents $\mathbf{z}_v$ and $\mathbf{z}_a$ as six contiguous frames within a unified tensor $\mathbf{P}$. This stacking strategy preserves local spatiotemporal continuity and allows the use of standard video diffusion transformers without architectural modifications.  
Second, this unified latent representation enables joint modeling of audio-video semantics through a single spatiotemporal attention mechanism, while maintaining computational efficiency.

\begin{figure}[t]
    \centering
    \includegraphics[width=0.9\linewidth]{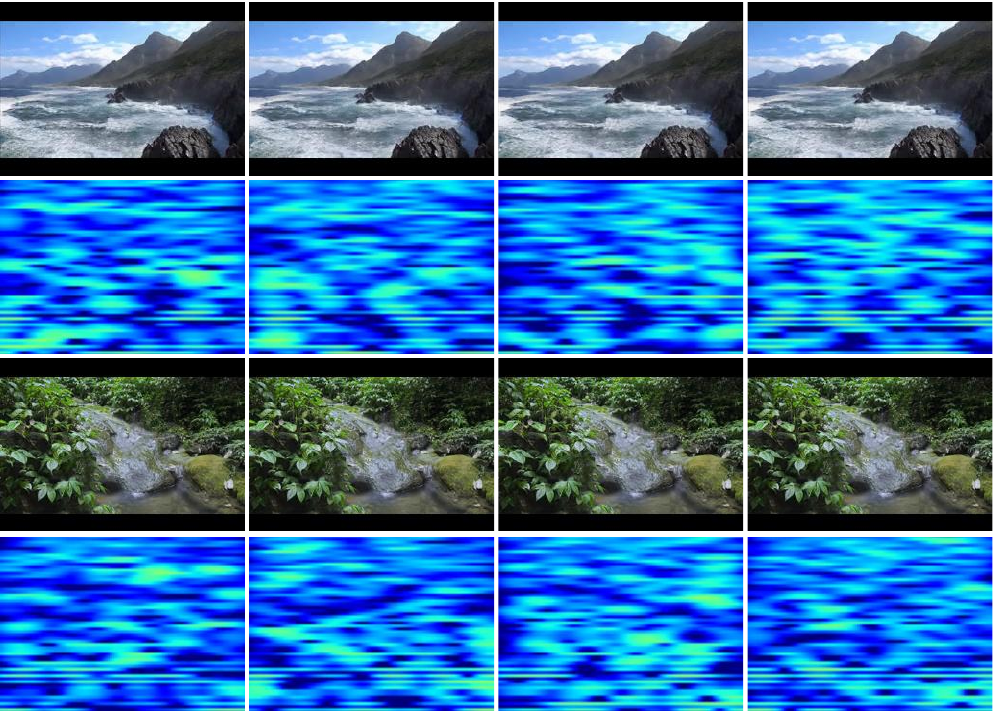}
    \caption{Results of our method on Landscape, including spectrogram visualization images and video frames.}
    \label{landscape-results}
    \vspace{-2ex}
\end{figure}

\begin{figure*}[t]
    \centering
    \includegraphics[width=0.9\linewidth]{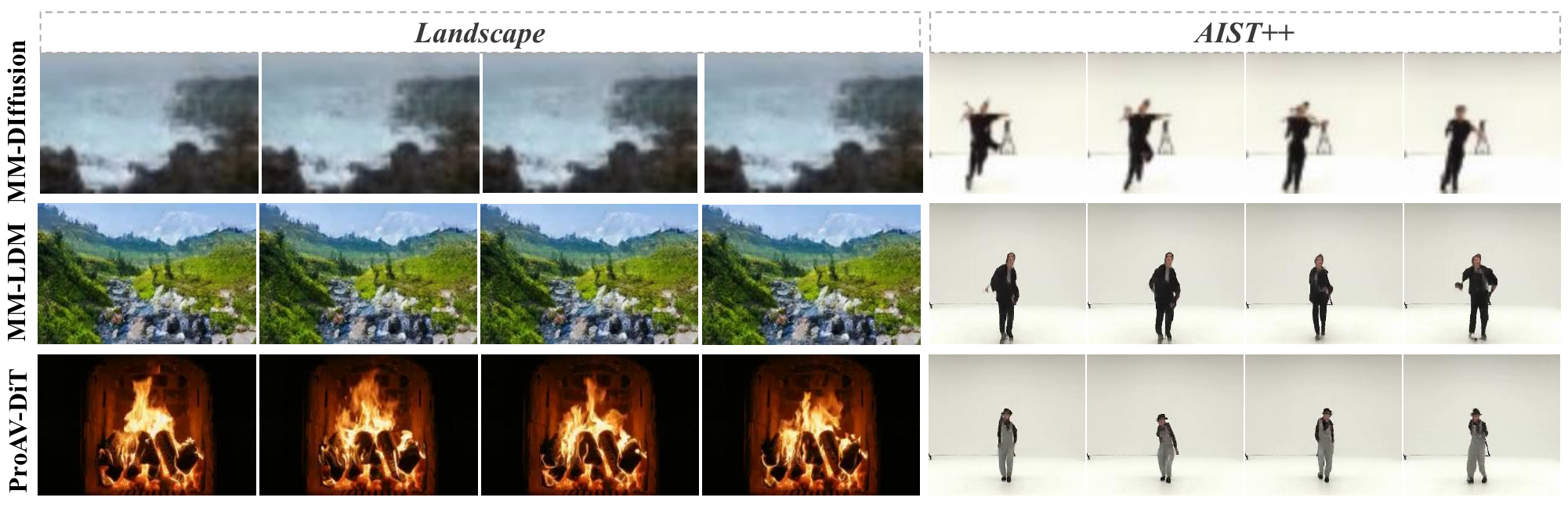}
    \caption{Qualitative comparison of ProAV-DiT with MM-Diffusion and MM-LDM.}
    \label{comparison-results}
    \vspace{-1ex}
\end{figure*}

\section{Experiment}
\subsection{Experimental Setups}
\paragraph{Datasets and Evaluation Metrics.} Following Ruan et al. \citep{ruan2023mm}, we evaluate our model on three datasets: Landscape \citep{lee2022sound}, AIST++ \citep{li2021ai}, and AudioSet \citep{7952261}. Both datasets are preprocessed into $16$-frame video clips with each frame resized to $256^2$ resolution. Details are provided in the supplementary material. We use Fréchet Video Distance (FVD) and Kernel Video Distance (KVD) to assess video quality, and Fréchet Audio Distance (FAD) to measure audio fidelity, aligning with prior studies \citep{sun2024mm, ruan2023mm, AV-DiT}. All videos generated by ProAV-DiT are synthesized at $256^2$ resolution.

\paragraph{Implementation Details.} 
The autoencoder and diffusion model are trained using the Adam \citep{adam} and AdamW \citep{adawm} optimizers, respectively. For audio preprocessing, we convert raw waveforms into mel-spectrograms with 128 mel bands.
The dual-stream encoder is based on Timesformer \citep{timesformer}, which serves as the backbone for projecting video into 3D latents. 
A two-stage training scheme is adopted. We first minimize perceptual loss, followed by adversarial loss and KL divergence loss ($\beta = 6\text{e}^{-6}$). For efficient learning and accelerated generation, we utilize the Reflected Flow Matching objective for training the diffusion model. We adopt STDiT as the generative model, and all reported results in this paper are obtained using 16 STDiT blocks. Video and audio discriminators are jointly trained with equal weighting (0.5). A linear noise schedule and reflected-flow sampling are used to accelerate inference. Details are provided in the supplementary material.
% A two-stage training strategy is employed. The initial stage involves minimizing the perceptual loss, which is subsequently followed by the simultaneous minimization of the adversarial loss and the KL divergence loss. The dual-stream encoder architecture is predicated upon the Timesformer \citep{timesformer} framework, which functions as the backbone for projecting video data into three-dimensional (3D) latent representations.
% To enhance learning efficiency and accelerate the generation process, the Reflected Flow Matching objective is utilized for training the diffusion model. STDiT is adopted as the foundational generative model. The video and audio discriminators are trained concurrently. Finally, a linear noise schedule and reflected-flow sampling are implemented to expedite the inference phase. Comprehensive details regarding optimization, specific parameter values, and data preprocessing are furnished in the supplementary material.

\begin{table*}[t]
\centering
% \small 
\caption{Quantitative performance comparison of multimodal video generation models on Landscape and AIST++ datasets. Results with $*$ are reproduced using released sources.}
\scalebox{0.95}{
% \begin{tabular}{l|c|c|ccc|ccc}
\begin{tabular}{@{}lccccccccc@{}}
\toprule
% \multirow{2}{*}{\textbf{Method}} & \multirow{2}{*}{\textbf{Resolution}} & \multirow{2}{*}{\textbf{Sampler}} & \multicolumn{3}{c|}{\textbf{Landscape}} & \multicolumn{3}{c}{\textbf{AIST++}} \\
% & & & FVD $\downarrow$ & KVD $\downarrow$ & FAD $\downarrow$ & FVD $\downarrow$ & KVD $\downarrow$ & FAD $\downarrow$ \\
\multirow{2}{*}{\textbf{Method}} & \multirow{2}{*}{\textbf{Resolution}} & \multirow{2}{*}{\textbf{Sampler}} & \multicolumn{3}{c}{\textbf{Landscape}} & \multicolumn{3}{c}{\textbf{AIST++}} \\
\cmidrule(lr){4-6} \cmidrule(lr){7-9} % 添加分组横线
& & & FVD $\downarrow$ & KVD $\downarrow$ & FAD $\downarrow$ & FVD $\downarrow$ & KVD $\downarrow$ & FAD $\downarrow$ \\
% \midrule
\midrule
% \multicolumn{9}{c}{\textbf{Single-Modal Video Generation}} \\
% \midrule
\multicolumn{9}{@{}c@{}}{\cellcolor{gray!30} Single-Modal Video Generation} \\ 
% \addlinespace[-2pt] 
% \midrule[0.5pt] 
\hdashline
DIGAN*              & 64$^2$ & -                 & 305.4 & 19.6  & -   & 119.5 & 35.8  & - \\
TATS-base*           & 64$^2$ & -                 & 600.3 & 51.5  & -   & 267.2 & 41.6  & - \\
MM-Diffusion-v*                  & 64$^2$ & dpm-solver        & 237.9 & 13.9  & -   & 163.1 & 28.9  & - \\
MM-Diffusion-v+SR*               & 64$^2$ & dpm-solver+DDIM   & 225.4 & 13.3  & -   & 142.9 & 24.9  & - \\
MM-LDM-v*                         & 64$^2$ & DDIM              & 122.1 & 6.4   & -   & 83.1  & 13.1  & - \\
MM-Diffusion-v+SR*               & 256$^2$ & dpm-solver+DDIM  & 347.9 & 27.8  & -   & 225.1 & 51.9  & - \\
MM-LDM-v*                         & 256$^2$ & DDIM             & 156.1 & 13.0  & -   & 120.9 & 26.5  & - \\
\tableLineBlue ProAV-DiT-v                        & 256$^2$ & Reflect Flow     & \textbf{90.9}  & \textbf{7.8} & -   & \textbf{84.2} & \textbf{13.3} & - \\
\midrule
\multicolumn{9}{@{}c@{}}{\cellcolor{gray!30} Multi-Modal Generation} \\ % 同样修复
% \addlinespace[-2pt]
% \midrule[0.5pt]
\hdashline
% \midrule
MM-Diffusion-svg+SR*            & 64$^2$ & dpm-solver+DDIM  & 211.2 & 12.6  & 9.9  & 137.4 & 24.2  & 12.3 \\
% MM-LDM-a2v*                       & 64$^2$ & DDIM             & 89.2  & 4.2   & -    & 71.0  & 10.8  & - \\
% MM-LDM-v2a*                       & -      & DDIM             & -     & -     & 9.2  & -     & -     & 10.2 \\
MM-LDM-svg*                       & 64$^2$ & DDIM             & 77.4 & 3.2 & 9.1 & 55.9 & 8.2 & 10.2 \\
% \midrule
\hdashline
MM-Diffusion-svg+SR*            & 256$^2$ & dpm-solver+DDIM  & 332.1 & 26.6  & 9.9  & 219.6 & 49.1  & 12.3 \\
% MM-LDM-a2v*                       & 256$^2$ & DDIM             & 123.1 & 10.4  & -    & 128.5 & 33.2  & - \\
MM-LDM-svg*                       & 256$^2$ & DDIM             & 105.0 & 8.3   & 9.1  & 105.0 & 27.9  & 10.2 \\
AV-DiT*                           & 256$^2$  & -                & 172.7 & 15.4  & 11.2 & 68.8  & 21.0  & 10.2 \\
% \midrule
\hdashline
\tableLineBlue ProAV-DiT w/o text         & 256$^2$ & Reflect Flow     & 87.3  & 7.7   & 8.7  & 85.6  & 19.8  & 9.8 \\
\tableLineBlue ProAV-DiT                 & 256$^2$ & Reflect Flow     & \textbf{80.3}  & \textbf{7.3} & \textbf{8.5}  & \textbf{77.6} & \textbf{18.2} & \textbf{9.4} \\
\midrule
\multicolumn{9}{@{}c@{}}{\cellcolor{gray!30} 200 Samples (Follow the See\&Hear experimental setup)} \\ % 同样修复
% \addlinespace[-2pt]
% \midrule[0.5pt]
\hdashline
See\&Hear*                        & 256$^2$ & -                & 326.2 & 9.2   & 12.7 & -     & -     & - \\
AV-DiT*                        & 256$^2$ & -                & 260.5 & 9.2 & 14.1 & -     & -     & - \\
 \tableLineBlue ProAV-DiT                          & 256$^2$ & Reflect Flow     & \textbf{240.3} & \textbf{9.0} & \textbf{12.2} & -     & -     & - \\
\bottomrule
\end{tabular}
}
\label{tab:comparison}
\vspace{-1ex}
\end{table*}

\subsection{Quantitative and Qualitative Comparison}
% % \paragraph{Autoencoder.} Evaluated on two datasets, it surpasses the single-stream baseline with gains of 1.40 dB PSNR and 14.6 FVD in video, and 0.85 dB PSNR and 0.5 FAD in audio—demonstrating the effectiveness of hierarchical attention enhanced dual-branch reconstruction. The detailed results are provided in the Appendix.
% \paragraph{Autoencoder.} Evaluated on two datasets, it surpasses the single-stream baseline with gains of 1.40 dB PSNR and 14.6 FVD in video, and 0.85 dB PSNR and 0.5 FAD in audio—demonstrating the effectiveness of hierarchical attention-enhanced dual-branch reconstruction. The detailed results are provided in the Fig.~\ref{tab:pvdm-comparion}.
% \paragraph{Qualitative Comparison.} Fig.~\ref{comparison-results} presents a qualitative comparison among ProAV-DiT, MM-LDM, and MM-Diffusion. Samples generated by MM-Diffusion exhibit noticeable blurriness and lack of detail, whereas MM-LDM produces clearer results with better audio-video alignment but still falls short of ProAV-DiT in realism and visual fidelity. Fig.~\ref{landscape-results} displays additional samples generated by ProAV-DiT on the Landscape dataset, further highlighting its superior generation quality. The autoencoder produces reconstructions that are visually indistinguishable from the ground truth, as detailed in the Appendix. We also present audio reconstruction results in the appendix, where the recovered waveforms closely match the originals, demonstrating the effectiveness of our heterogeneous modality unification.

\paragraph{Qualitative Comparison.} Fig.~\ref{comparison-results} shows a qualitative comparison among ProAV-DiT, MM-LDM, and MM-Diffusion. MM-Diffusion produces blurry, low-detail samples, while MM-LDM achieves clearer outputs with improved audio-video alignment but still lags behind ProAV-DiT in realism and fidelity. Fig.~\ref{landscape-results} further illustrates ProAV-DiT’s superior generation quality on the Landscape dataset. Moreover, the autoencoder yields reconstructions visually indistinguishable from ground truth, and our audio reconstruction results (Appendix) demonstrate that recovered waveforms closely align with the originals, validating the effectiveness of our heterogeneous modality unification.

\paragraph{Performance Comparison with Previous Methods.} 
We performed a quantitative comparison of the ProAV-DiT method against existing approaches to validate its effectiveness. As illustrated in Table~\ref{tab:comparison}, when utilizing conditional input, ProAV-DiT achieved an average improvement of $7.3\%$ in $\text{FVD}$ on the Landscape dataset and an average improvement of $8.0\%$ in $\text{FVD}$ on the AIST++ dataset (relative to unconditional generation).
ProAV-DiT employs an end-to-end training framework at $256^2$ resolution, where the autoencoder directly processes the video at the original resolution. This design generates high-quality output without requiring additional super-resolution modules. Specifically, ProAV-DiT achieves competitive results with 80.3\% FVD, 7.3\% KVD, and 8.5\% FAD on the Landscape dataset, and 77.6\% FVD, 18.2\% KVD, and 9.4\% FAD on the AIST++ dataset. This demonstrates its superior ability to capture cross-modal correlations. These performance improvements are primarily attributed to the dual-stream encoder, which optimizes the alignment of cross-modal representations, thereby improving the fidelity of the generated video and audio content. 
To evaluate the generalization and scalability of ProAV-DiT, we conducted experiments on a larger open-domain dataset. Following the protocol established by MM-Diffusion, we selected $100\text{k}$ high-quality videos from the AudioSet. As indicated in Table~\ref{tab:scale}, ProAV-DiT consistently outperforms previous methods while utilizing fewer parameters, thereby substantiating its effectiveness and scalability on large datasets.

\vspace{-2ex}
\paragraph{Efficiency Comparison.} 
Unlike MM-Diffusion, which operates in the signal space and encounters memory limitations at $256^2$ resolution, ProAV-DiT efficiently handles $256^2$ input by leveraging the MDSA autoencoder. While both MM-LDM and ProAV-DiT integrate an autoencoder with the Diffusion Transformer (DiT), a comparison isolating the DiT performance (MM-LDM* and ProAV-DiT*, achieved by pre-computing and storing latents) reveals that ProAV-DiT* further boosts efficiency. At a batch size of 2, ProAV-DiT trains at $0.31$s per step, which is faster than MM-LDM ($0.38$s) and substantially faster than MM-Diffusion ($2.36$s at the lower $128^2$ resolution). For inference, Reflected Flow Sampling \citep{xie2024reflectedflowmatching} reduces sampling steps from 100–200 to 30, cutting runtime to $3.9s$ per sample, yielding 2.2× and 22.5× speedups over MM-LDM and MM-Diffusion. 

\begin{table}[t]
    \centering
        \centering
        \caption{Quantitative comparison on the Audioset for open-domain generation.}
        \vspace{-1ex}
        \label{tab:scale}
        \scalebox{0.9}{
        \begin{tabular}{l c c c c}
        \toprule
        \textbf{Model} & \textbf{\#P} & 
        \multicolumn{1}{c}{\textbf{FVD}$\downarrow$} & 
        \multicolumn{1}{c}{\textbf{KVD}$\downarrow$} & 
        \multicolumn{1}{c}{\textbf{FAD}$\downarrow$} \\
        \midrule
        MM-Diffusion   & 134M    & 649.8 & 34.6 & 2.9 \\
        MM-LDM-S       & 131M    & 185.8 & 10.1 & 1.59 \\
        MM-LDM-B       & 384M    & 181.5 & 9.5  & 1.55 \\
        MM-LDM-L       & 1543M   & 164.1 & 8.5  & 1.52 \\
       \tableLineBlue ProAV-DiT        & 702M    & 148.7 & 8.4  & 1.51 \\
        \bottomrule
        \end{tabular}
        }
        \vspace{-2ex}
\end{table}

% \subsection{AV-Alignment and User Study}
% \input{table/tab3}
% To quantify the degree of audio-visual alignment, we introduce the objective metrics CLAP-Similarity (CS) and AV-Align \cite{yariv2024diverse} (details provided in Appendix~\ref{av-metrics}), applying them to 1,500 samples generated by ProAV-DiT. Specifically, CS measures semantic consistency, whereas AV-Align quantifies temporal coherence. Our ProAV-DiT model demonstrated superior performance, achieving a CS score of 0.77 and an AV-Align score of 0.63, which confirms its enhanced capability in capturing cross-modal correlations. We also conducted a subjective human evaluation, following the MM-Diffusion protocol, using the Landscape dataset to compare ProAV-DiT against MM-Diffusion and MM-LDM. Two annotators rated each sample on a 5-point Likert scale across three core criteria: Audio Quality (AQ), Video Quality (VQ), and Audio-Video Alignment (A-V). As summarized in Table~\ref{human}, ProAV-DiT consistently outperforms the baselines, registering relative gains over MM-LDM of 8.1\% in AQ, 7.9\% in VQ, and 9.4\% in A-V.

\subsection{AV-Alignment and User Study}

\begin{table}[t]
        \centering
        \caption{Efficiency comparison of ProAV-DiT with other methods.}
         % \vspace{-1ex}
        \label{tab:efficiency}
        \scalebox{0.9}{
        \begin{tabular}{@{}l c >{\centering\arraybackslash}p{1.8cm} >{\centering\arraybackslash}p{1.8cm}@{}}
        \toprule
        \textbf{Method} & \textbf{Res.} & \textbf{Train/Step} & \textbf{Infer/Sample} \\
        \midrule
        MM-Diffusion          & $64^2$    & 1.70s  & 33.8s \\
        MM-Diffusion          & $128^2$   & 2.36s  & 90.0s \\
        % \midrule
        MM-LDM         & $256^2$   & 0.46s  & 70.0s \\
        MM-LDM*        & $256^2$   & 0.38s  & 8.7s  \\
        % \midrule
       \tableLineBlue ProAV-DiT (ours)        & $256^2$   & 0.44s  & 17.7s \\
       \tableLineBlue ProAV-DiT* (ours)       & $256^2$   & \textbf{0.32s}  & \textbf{3.93s} \\
        \bottomrule
        \end{tabular}
        % }
    % \vspace{-3ex}
    }
    \vspace{-3ex}
\end{table}

To quantify the degree of audio-visual alignment, we introduce two objective metrics, namely CLAP-Similarity (CS) and AV-Align \cite{yariv2024diverse} (detailed in Appendix~\ref{av-metrics}) on 1,500 samples generated by ProAV-DiT. Specifically, CS measures semantic consistency, whereas AV-Align quantifies temporal coherence. Our proposed ProAV-DiT model demonstrate superior performance on both metrics, achieving a CS score of $0.306$ and an AV-Align score of $0.63$. Furthermore, we conducted a subjective human evaluation following the protocol established by MM-Diffusion. This study utilized the Landscape dataset to compare ProAV-DiT against established baselines, namely MM-Diffusion and MM-LDM, as well as See\&Hear and AV-DiT (for AV-Align). Two independent annotators rated each generated sample on a 5-point Likert scale across three core quality criteria: Audio Quality (AQ), Video Quality (VQ), and Audio-Video Alignment (A-V). As summarized in Table~\ref{human}, ProAV-DiT consistently outperforms all baselines. Notably, compared to the strong baseline MM-LDM, ProAV-DiT registered significant relative gains of $8.1\%$ in AQ, $7.9\%$ in VQ, and $9.4\%$ in A-V. The high AV-Align score of $0.63$ also confirms its state-of-the-art performance in temporal synchronization.

\begin{table}[t]
 \centering
 \caption{Audio-video alignment results and human evaluation.}
 \scalebox{0.85}{ % 缩放为原大小的85%
 \begin{tabular}{lccc|cc}
 \toprule
 \textbf{Method} & \textbf{AQ}$\uparrow$ & \textbf{VQ}$\uparrow$ & \textbf{A-V}$\uparrow$ &  \textbf{CS}$\uparrow$ &  \textbf{AV-Align}$\uparrow$\\
 \midrule
 
 MM-Diffusion & 2.46 & 2.10 & 2.99 & 0.268 & 0.53\\
 MM-LDM & 2.98 & 3.68 & 3.29 & 0.285 & 0.57\\
 See\&Hear & - & - & - & - & 0.57 \\
 AV-DiT & - & - & - & - & 0.59 \\
 % \midrule
\tableLineBlue ProAV-DiT& 3.22 & 3.97 & 3.60 & 0.306 & 0.63\\
 \bottomrule
 \end{tabular}
 }
 \label{human}
 \vspace{-3ex}
\end{table}

\subsection{Ablation Study}
We conducted an ablation study on the architecture of MDSA, with the results summarized in Table~\ref{tab:unified-mdsa-results-merged-fvd}. 
The upper half of Table~\ref{tab:unified-mdsa-results-merged-fvd} presents the reconstruction performance of different configurations. The MDSA module demonstrates strong performance, with values of $\text{rFVD}$ of 18.7, $\text{rFAD}$ of 8.7, and $\text{PSNR}$ of 32.19. Additionally, we evaluate two single-stream orthogonal VAE configurations, namely MDSA-V and MDSA-A. MDSA-V achieves $\text{rFVD}$ of 30.3, while MDSA-A achieves $\text{rFAD}$ of 9.4. These results further emphasize the importance of cross-modal alignment, as MDSA outperforms its single-stream counterparts, highlighting the advantages of unified encoding.

The lower half of Table~\ref{tab:unified-mdsa-results-merged-fvd} focuses on the generation quality of MDSA.
The base MDSA configuration achieves superior performance with $\text{rFVD}$ of 29.5,  $\text{rFAD}$ of 8.5 and $\text{rKVD}$ of 1.3. These results demonstrate that MDSA not only excels in reconstruction but also in generating high-quality synchronized audio-video content. To evaluate the impact of the multi-scale attention mechanism, we sequentially remove MT-selfAttn, GCM-Attn, and Bi-Block CrossAttn, which leads to increased $\text{rFVD}$ scores of 35.1, 39.2, and 45.6, respectively, confirming the effectiveness of our attention mechanism design. Eliminating KL regularization leads to degraded performance, with $\text{rFVD}$ increasing to 60.2, which demonstrates the benefit of regularization-based fine-tuning. 
%Furthermore, to rigorously validate the superiority of MDSA, we conducted a quantitative comparison of its reconstruction capability against standard Variational Autoencoder (MM-LDM) architectures. The quantitative results, presented in Table~\ref{tab:unified-mdsa-results-merged-fvd}, clearly indicate that MDSA achieves notably lower $\text{rFVD}$ and $\text{rFAD}$ scores. The performance advantage over conventional VAEs further demonstrates the role of the unified latent space in promoting modality synergy. MDSA-V and MDSA-A represent the single-stream orthogonal VAE configurations. The significant performance gains exhibited by MDSA over these single-stream orthogonal VAEs underscore the positive impact of cross-modal alignment.

% \input{table/tab5}
\begin{table}[t]
\centering
\caption{Ablation study on the MDSA architecture.}
\label{tab:unified-mdsa-results-merged-fvd}
\scalebox{0.8}{
\begin{tabular}{lccccc}
\toprule
\textbf{Method/Setting} & \textbf{rFVD}$\downarrow$ & \textbf{rFAD}$\downarrow$ & \textbf{rKVD}$\downarrow$ & \textbf{PSNR}$\uparrow$ \\
\midrule
\multicolumn{5}{c}{\cellcolor{gray!15}\textbf{Reconstruction Performance (AIST++ Dataset)}} \\
\midrule
MM-LDM & 53.9 & 8.9 & - & - \\
MDSA-V & 30.3 & - & - & 31.34 \\
MDSA-A & - & 9.4 & - & - \\
\tableLineBlue \textbf{MDSA} & \textbf{18.7} & \textbf{8.7} & - & \textbf{32.19} \\
\midrule
\multicolumn{5}{c}{\cellcolor{gray!15}\textbf{Ablation Study on Generation Quality (Landscape Dataset)}} \\
\midrule
\tableLineBlue \textbf{MDSA} & \textbf{29.5} & 8.5 & \textbf{1.3} & - \\
% \midrule
$-$ MT-selfAttn & 35.1 & 8.6 & 2.2 & - \\
$-$ GCM-Attn & 39.2 & 8.7 & 2.3 & - \\
$-$ Bi-Block CrossAttn & 45.6 & 8.8 & 2.4 & - \\
% \midrule
$-$ finetune w/o KL loss & 60.2 & 8.9 & 3.2 & - \\
$-$ w/o adversarial loss & 134.5 & 9.8 & 8.3 & - \\
\bottomrule
\end{tabular}
}
\vspace{-3ex}
\end{table}

\vspace{-1ex}
\section{Conclusion}
We introduce ProAV-DiT, a novel diffusion transformer designed for the SVG task.
We propose a unified projection autoencoder that maps both audio and video into a shared latent space by projecting 3D data into a 2D representation. Furthermore, the use of a multi-scale dual-stream spatiotemporal autoencoder and the multi-scale attention mechanism strengthens temporal synchronization and bridges the semantic gap between modalities. Built upon the STDiT architecture, ProAV-DiT enables rich cross-modal interactions during the generation process.
Our method achieves new state-of-the-art results across multiple benchmarks, demonstrating superior efficiency and promising adaptability.

{
    \small
    \bibliographystyle{ieeenat_fullname}
    \bibliography{main}
}

\clearpage
\setcounter{page}{1}
\maketitlesupplementary

\subsection{Datasets}
\paragraph{Landscape Dataset.} The Landscape dataset focuses on high-fidelity audiovisual synchronization of natural scenes. It comprises 928 landscape videos crawled from YouTube, covering nine representative natural scenarios such as rainfall, splashing water, thunderstorms, and underwater bubbles. Each video clip is annotated with scene category, weather conditions, and types of acoustic events (e.g., “heavy rain + thunder”), supporting fine-grained conditional generation tasks. After preprocessing, the dataset yields 1,000 non-overlapping 10-second video clips, with a total duration of approximately 2.7 hours. The audio tracks feature a high dynamic range of environmental sounds that are tightly aligned with the visual scenes (e.g., thunder sounds coinciding with lightning flashes), providing naturally aligned annotations for cross-modal learning tasks.

\paragraph{AIST++ Dataset.}AIST++ is constructed based on the AIST street dance database and consists of 1,020 dance video clips (with a total duration of 5.2 hours), accompanied by 60 copyright-free music tracks spanning 10 dance genres (e.g., Hip-Hop, Krump, Ballet Jazz). The dataset includes 85\% basic choreography and 15\% freestyle movements, enhancing the model's ability to adapt to diverse musical styles. It provides 9 camera pose parameters, 17 COCO-format 2D/3D keypoints, 24-dimensional SMPL pose parameters, and global motion trajectories. Its core value lies in the precise spatiotemporal alignment between dance movements and music. Through multi-view camera calibration and SMPL-based 3D human motion reconstruction, the dataset offers 3D motion sequences with joint rotations and displacement information, along with annotated music beat timestamps.

\subsection{Implementation Details}
\paragraph{Detailed description of training objective}
% \section{Training Objective}
The proposed multi-modal adversarial training objective jointly optimizes reconstruction constraints and distribution alignment through a dynamically scheduled optimization framework. The overall loss function is defined as:
\begin{equation}
\mathcal{L}_{\text{total}} = \sum_{m \in \{\text{video}, \text{audio}\}} \omega_m \left( \mathcal{L}_{\text{rec}} + \mathcal{L}_{\text{perc}} + \gamma(\mathcal{L}_{\text{adv}}^{(G)} + \mathcal{L}_{\text{fm}}) \right)
\end{equation}
where $\omega_m = 0.5$ balances cross-modal weights and $\gamma$ denotes the dynamic adversarial activation factor. Specifically, the multimodal reconstruction loss combines pixel-level fidelity with latent space regularization:
\begin{equation}
\mathcal{L}_{\text{rec}} = \underbrace{4.0 \cdot \| \mathbf{x} - \hat{\mathbf{x}} \|_1}_{\text{pixel fidelity}} + \underbrace{6 \times 10^{-6} \cdot D_{\text{KL}} \left( \mathcal{Q}(\hat{\mathbf{x}}) \parallel \mathcal{P}(\mathbf{x}) \right)}_{\text{distribution alignment}}
\end{equation}
with KL-divergence computed using batchmean reduction: $\text{KL}(p \parallel q) = \sum p(x) \log \frac{p(x)}{q(x)}$. 

We adopt a two-stage training strategy to stabilize optimization. During the first stage, both the adversarial loss $\mathcal{L}_{\text{adv}}^{(G)}$ and KL regularization are disabled, allowing the model to focus on basic reconstruction. In the second stage, we enable adversarial training and KL divergence to enhance visual fidelity and latent alignment.
The dynamic adversarial activation is governed by:
\begin{equation}
\label{eq:gamma}
\gamma = 
\begin{cases}
0 & \text{if } t < t_{\text{threshold}} \\
1.0 & \text{otherwise}
\end{cases}
\end{equation}
where $t_{\text{threshold}}$ denotes the training step at which the discriminator becomes active (controlled via the \texttt{disc\_start} parameter in code).

\paragraph{Training Details.} 
For all experiments, we use a batch size of 32 and a learning rate of $1 \times 10^{-4}$ to train the autoencoders. Training continues until both FVD and PSNR metrics converge. For 3D-to-2D projection, we employ a 4-layer Transformer with 4 attention heads, a hidden dimension of 384, and an MLP dimension of 512. The latent codebook dimensionality is set to 4.
For diffusion model training, we use a batch size of 64 and the same learning rate of $1 \times 10^{-4}$. Additional architectural hyperparameters, we basically followed the parameters of Opensora \cite{sora}, but we used 16 layers in STDiT. Specifically, we set the codebook channel $C=4$ and the patch size to $4 \times 4 \times 1$, such that a video of size $256 \times 256 \times 16 \times 3$ is encoded into a latent vector of size $(32 \times 32 + 32 \times 16 + 32 \times 16) \times 4 = 8192$.

\paragraph{Metric.}
To ensure a fair comparison with prior work, we adopt consistent settings for quantitative evaluation. For Fréchet Video Distance (FVD) and Fréchet Audio Distance (FVD), we follow the fixed protocol proposed by StyleGAN-V \cite{styleganv}. Unlike the standard protocol—which first preprocesses the dataset into fixed-length video clips before computing real statistics—the StyleGAN-V protocol samples video data first, then randomly extracts fixed-length clips. This adjustment addresses bias introduced when long videos dominate the dataset, skewing the statistics due to their excessive number of clips. 
Following MM-Diffusion, we sample 2,048 videos (or the full dataset if it contains fewer samples) to compute the real distribution and another 2,048 videos to evaluate the generated samples.

\begin{figure}[h]
    \centering
    \includegraphics[width=1.0\linewidth]{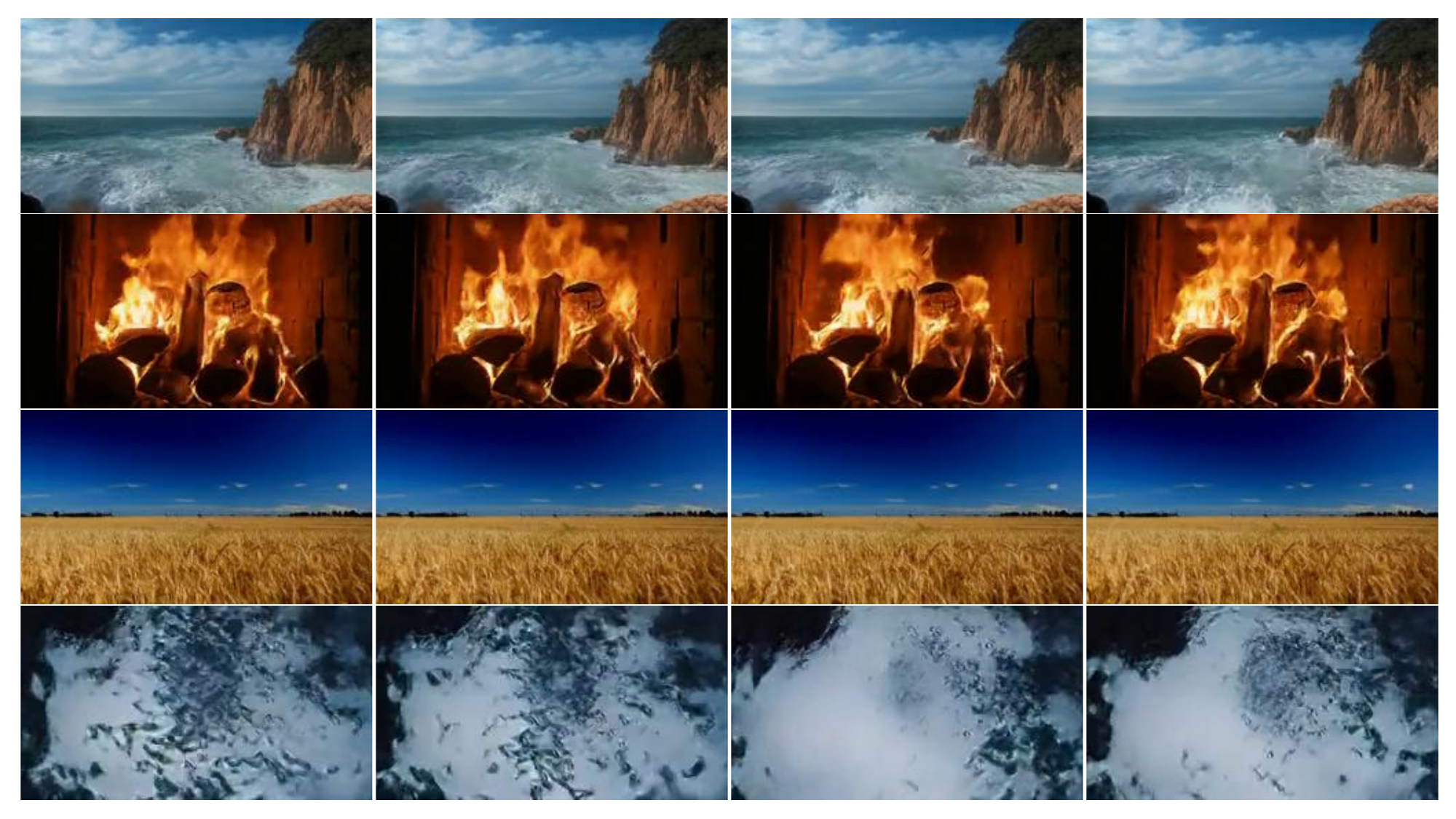}
    \caption{Video reconstruction results of our MDSA on the Landscape dataset}
    \label{landscape1Reconstruction}
\end{figure}

\begin{figure}[h]
    \centering
    \includegraphics[width=1.0\linewidth]{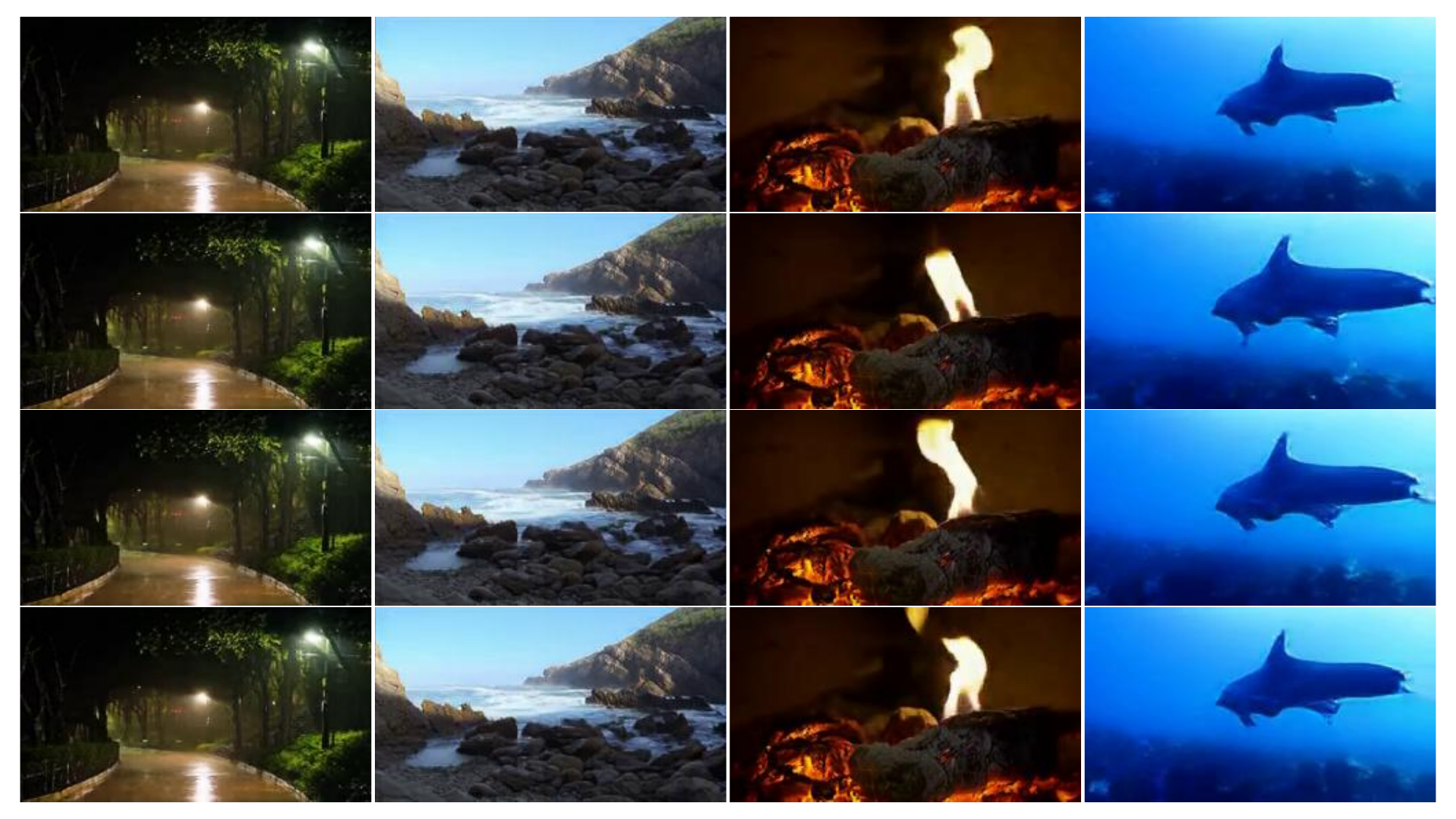}
    \caption{Video reconstruction results of our MDSA on the Landscape dataset}
    \label{landscape2Reconstruction}
\end{figure}

\begin{figure}[h]
    \centering
    \includegraphics[width=1.0\linewidth]{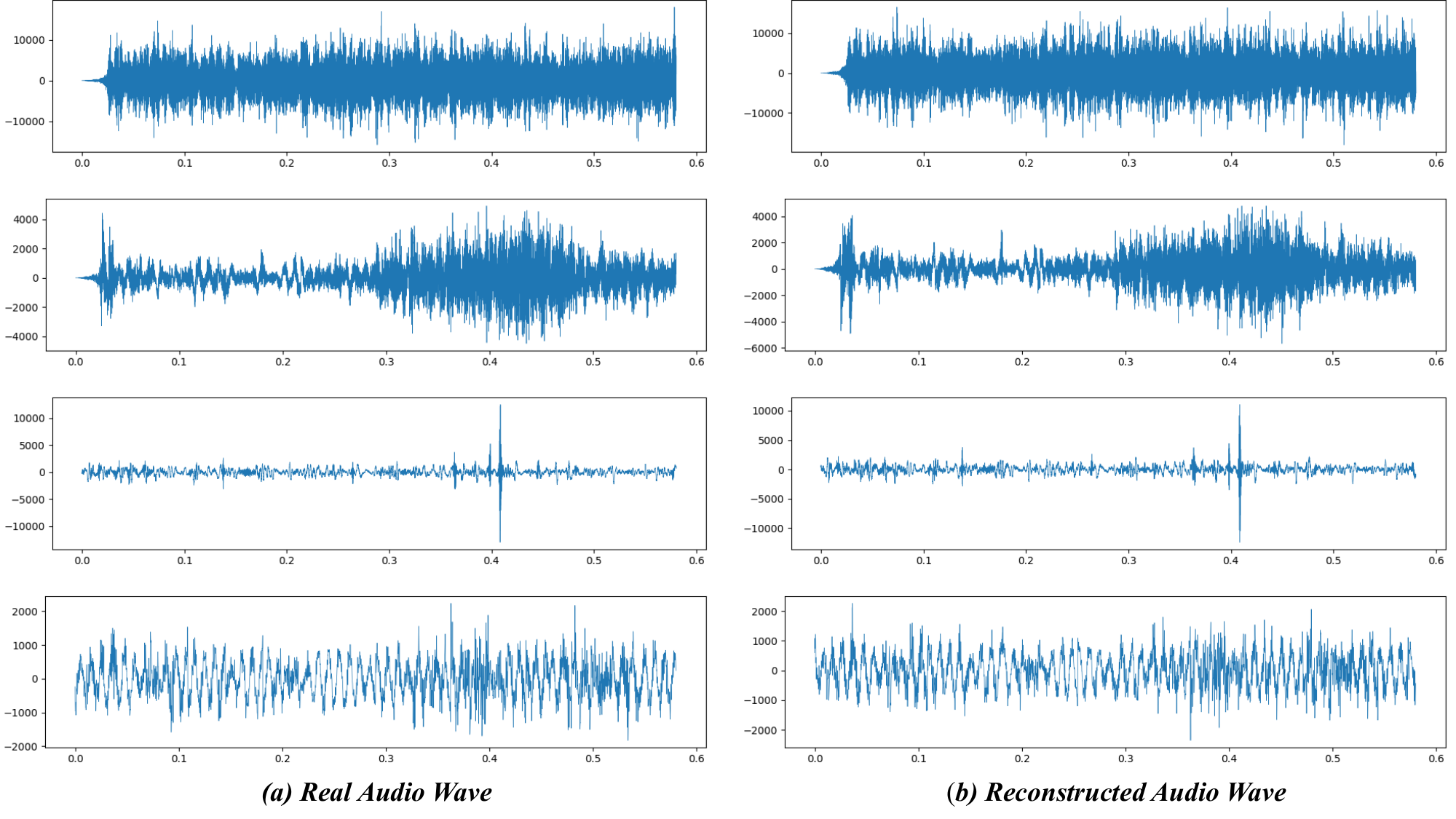}
    \caption{Audio reconstruction results of our MDSA on the Landscape dataset}
    \label{audio-Reconstruction}
\end{figure}

\begin{figure}[h]
    \centering
    \includegraphics[width=1.0\linewidth]{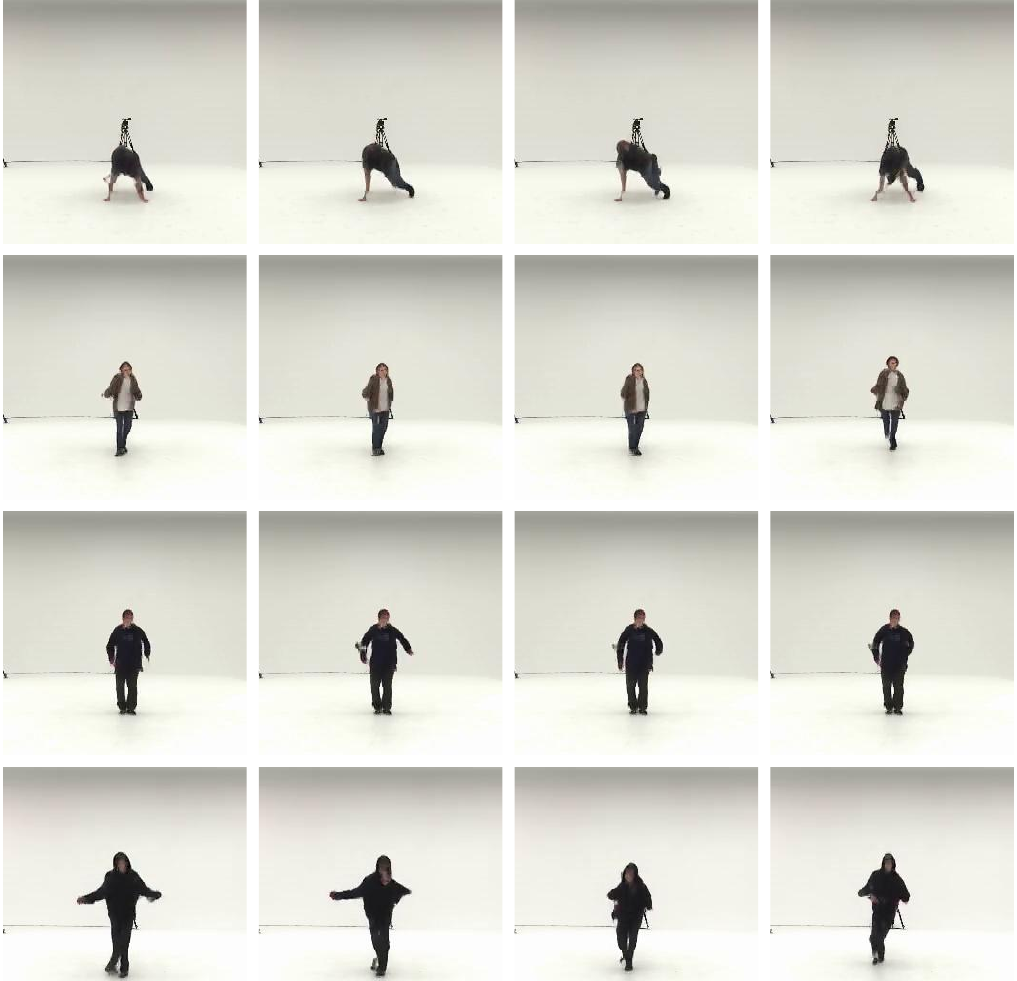}
    \caption{Results of our method on AIST++.}
    \label{aist}
\end{figure}
\subsection{Quantitative Comparison}
\begin{table}[t]
\centering
% \vspace{-3ex} % 保持排版简洁，去除不必要的硬编码垂直间距
\caption{Comparison of MDSA Performance (Reconstruction) on the Landscape Dataset.} % 优化标题以更清晰地说明是重建性能
\label{tab:pvdm-comparion}
\begin{tabular}{lcccc} 
\toprule
\textbf{Method} & \textbf{FVD}$\downarrow$ & \textbf{FAD}$\downarrow$ & \textbf{PSNR-A}$\uparrow$ & \textbf{PSNR-V}$\uparrow$ \\ % 统一指标方向
\midrule
\tableLineBlue \textbf{A-V Rec.} & \textbf{18.7} & \textbf{8.7} & \textbf{37.33} & \textbf{32.19} \\
% \midrule
% Ablation for clarity (Optional, based on original content)
V Rec. & 30.3 & - & - & 31.34 \\ % 仅视频重建（可能用于消融对比）
A Rec. & - & 9.4 & 35.93 & - \\ % 仅音频重建（可能用于消融对比）
\midrule
MM-LDM  & 53.9 & 8.9 & - & -  \\
\bottomrule
\end{tabular}
\end{table}

\paragraph{Autoencoder Performance.}
\label{AE}
On the Landscape dataset, we systematically compared the reconstruction performance of different autoencoders to validate the superiority of the proposed Multi-scale Dual-stream Spatio-temporal Autoencoder (MDSA). As shown in Table~\ref{tab:pvdm-comparion}, the evaluation encompassed four methods: MM-LDM served as the conventional VAE baseline; V Rec. and A Rec. represented the single-stream orthogonal VAEs under the MDSA framework (i.e., video or audio processed separately); while A-V Rec. denoted the full MDSA (modality-interactive version), which employs cross-modal attention for joint encoding and reconstruction.
Specifically, MDSA demonstrated excellent performance across all metrics. In video reconstruction, it achieved an FVD of 18.7, significantly lower than the 30.3 of the single-stream video VAE (V Rec.) and 53.9 of the conventional VAE (MM-LDM); simultaneously, it attained a PSNR-V of 32.19 dB, higher than the 31.34 dB of V Rec., indicating that the interactive design enhances visual quality. In audio reconstruction, MDSA achieved an FAD of 8.7, outperforming the 9.4 of the single-stream audio VAE (A Rec.) and 8.9 of MM-LDM; its PSNR-A of 37.33 dB also exceeded the 35.93 dB of A Rec., highlighting the positive impact of cross-modal alignment.
Overall, compared to single-stream VAEs, MDSA reduced FVD by 11.6 (a relative improvement of 38.2\%) and increased PSNR-V by 0.85 dB in video; for audio, it reduced FAD by 0.7 (a relative improvement of 7.4\%) and increased PSNR-A by 1.40 dB. These quantitative results validate the effectiveness of MDSA in reducing redundancy and enhancing synchronization through orthogonal decomposition and multi-scale attention mechanisms. The performance advantage over conventional VAEs further demonstrates the role of the unified latent space in promoting modality synergy.
The detailed results are provided in the Fig.~\ref{tab:pvdm-comparion}.

\subsection{MI}
\label{MI}
To quantify the degree of decoupling in our proposed feature decomposition ($z_{xy}$, $z_{yt}$, and $z_{xt}$, representing spatial, vertical-temporal, and horizontal-temporal features, respectively), we calculated the pairwise Mutual Information (MI) between these latent representations. From an information-theoretic perspective, the theoretical basis of our method lies in minimizing cross-axis redundancy, meaning we aim for the values of $MI(z_{xy}; z_{yt})$, $MI(z_{xy}; z_{xt})$, and $MI(z_{yt}; z_{xt})$ to be as low as possible. 
\begin{table}[htbp]
\centering
\caption{Mutual Information Feature Association Analysis}
\label{tab:mi_analysis}
\begin{tabular}{l l c c l}
\toprule
MI Pair  & Mean & Std & Decoupling Degree \\
\midrule
XY-YT &  0.3204 & 0.1388 & Weakest \\
XY-XT &  0.252 & 0.1328 & Strongest \\
YT-XT &  0.306 & 0.1347 & Intermediate \\
\bottomrule
\end{tabular}
\end{table}
The Table~\ref{tab:mi_analysis} reports the statistical results of the MI estimates collected during training (measured in nats). The results indicate that $MI_{xy\_xt}$ has the lowest average value, suggesting minimum redundancy and the best decoupling effect between the spatial features and the horizontal-temporal features. Conversely, $MI_{xy\_yt}$ exhibits the highest average value, which may be due to both spatial features (XY) and vertical-temporal features (YT) relying heavily on information from the vertical (Y) dimension, leading to relatively high information sharing and redundancy. Overall, these MI values provide quantitative evidence supporting that our decomposition structure achieves an effective degree of cross-axis information decoupling, thereby providing an information-theoretic basis for efficient and fine-grained cross-modal alignment.

\begin{figure*}[h]
    \centering
    \includegraphics[width=1.0\linewidth]{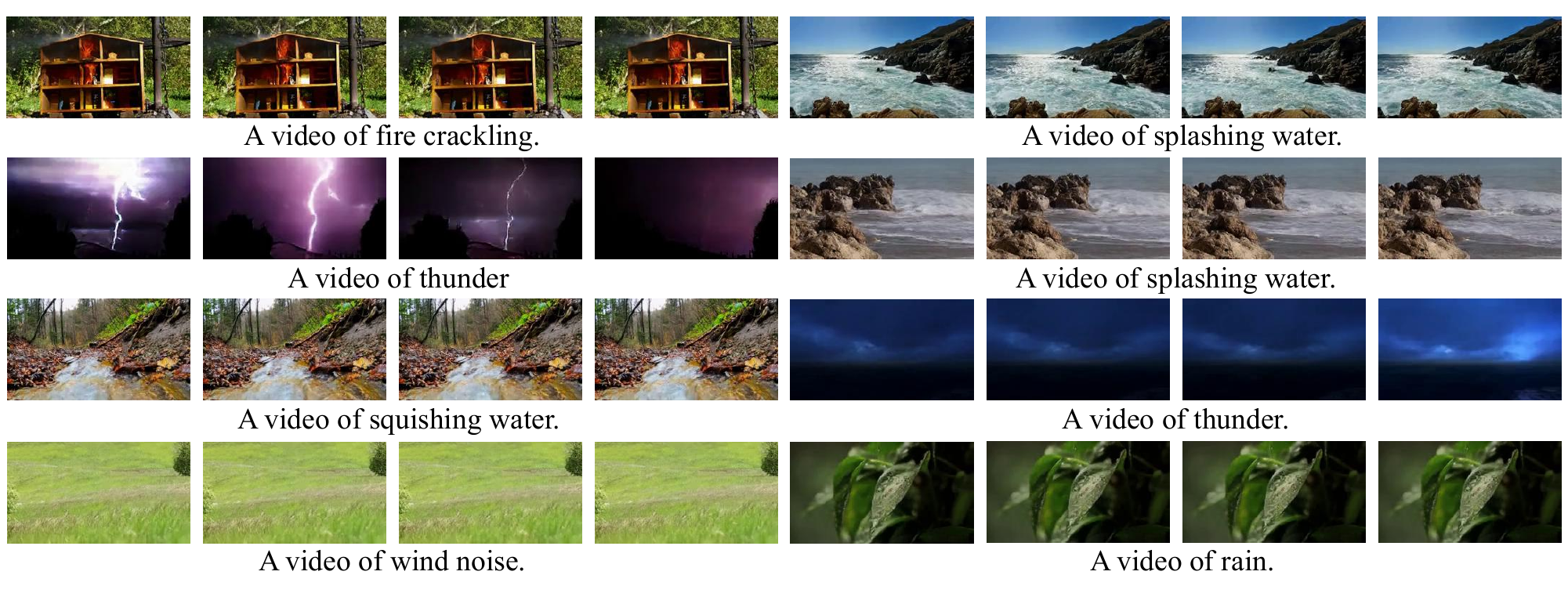}
    \caption{Our method's text-conditional guided generation results on the landscape dataset.}
    \label{text2video-landscape}
\end{figure*}

\subsection{AV-Alignment Metrics}
\label{av-metrics}
\paragraph{CLAP-Similarity.}
We introduce the objective CLAP-Similarity (CS) metric to measure audio-visual alignment. CS uses the CLAP mode to project audio and video into a shared latent space. Alignment is quantified by the cosine similarity between their embeddings. The CLAP encoders $E_a$ and $E_v$ extract feature vectors ($\mathbf{z}_{a, i}$ and $\mathbf{z}_{v, i}$) for each segment $i$. The final CS score is the average cosine similarity across all segments:$$\text{CS} = \frac{1}{W} \sum_{i=1}^{W} \text{CosineSim}(\mathbf{z}_{a, i}, \mathbf{z}_{v, i})$$A higher CS value indicates better semantic alignment. 

\paragraph{AV-Align.}
AV-Align is a sophisticated metric designed to directly measure the temporal coherence between audio and video by leveraging the fundamental premise that rapid energy changes in the audio signal (peaks) should correspond to the motion of the sound-producing object in the video.The operation flow of AV-Align is detailed as follows:

A. Event Peak DetectionWe detect moments of significant event occurrence from both modalities:Audio Peaks: An onset detection algorithm is used to precisely locate moments where the audio signal intensity abruptly increases (e.g., the start of a drum beat or knock).Video Peaks: Peaks are identified by calculating the mean magnitude of Optical Flow between consecutive frames. A high magnitude indicates a rapid motion, such as a drumstick striking a drum, marking a significant motion change.

B. Peak MatchingThe algorithm then checks whether each detected peak in one modality has a matching peak in the corresponding time window of the other modality.Matching Rule: For an audio peak, the system searches for a video peak within a three-frame time window before and after the audio peak's occurrence, and vice-versa. A match is recorded if a corresponding peak is found within this window.

C. The AV-Align score is calculated as the average of two directional peak matching success rates.The formula explicitly measures the proportion of audio peaks that successfully find a match in the video ($\frac{\sum_{a \in A} \mathbb{I}_{a \leftrightarrow V}}{|A|}$) and the proportion of video peaks that successfully find a match in the audio ($\frac{\sum_{v \in V} \mathbb{I}_{v \leftrightarrow A}}{|V|}$), and then averages these two ratios:$$\text{AV-Align} = \frac{1}{2} \left[ \frac{\sum_{a \in A} \mathbb{I}_{a \leftrightarrow V}}{|A|} + \frac{\sum_{v \in V} \mathbb{I}_{v \leftrightarrow A}}{|V|} \right]$$Here, $A$ and $V$ are the sets of audio and video peaks, and $\mathbb{I}_{\cdot \leftrightarrow \cdot}$ is the indicator function for a successful match within the defined temporal window. This approach averages the matching success rate in both directions (audio-to-video and video-to-audio), yielding a comprehensive score between 0 and 1. A higher score signifies better temporal synchronization between the audio and video on the timeline.

\subsection{Qualitative results}
\paragraph{Autoencoder.} Fig.~\ref{landscape1Reconstruction} and Fig.~\ref{landscape2Reconstruction} shows the results of our MDSA reconstruction of the Landscape dataset. As can be seen, our MDSA produces high-quality synthetic results overall. The precise reconstruction of the layered rock structure of the coastal cliffs, the instantaneous shape of the splashing waves and their gradual transition back to the sea surface, and the natural transition from the bright flame core to the orange outer flame all demonstrate the excellent performance of our autoencoder.
Figure~\ref{audio-Reconstruction} illustrates the audio reconstruction performance of our method. Subfigure (a) shows the ground-truth audio waveform, while subfigure (b) depicts the reconstructed waveform obtained after encoding and decoding via our autoencoder, followed by inverse Mel-spectrogram transformation. The close similarity between the two waveforms demonstrates that our autoencoder effectively preserves high-fidelity audio content. This accurate reconstruction of both audio and video provides a solid foundation for the subsequent high-fidelity generation within the diffusion model.

\paragraph{Diffusion generator.}
We present qualitative results of ProAV-DiT in Fig.~\ref{aist} and Fig.~\ref{text2video-landscape}, showcasing generated samples on the AIST++ and Landscape datasets, respectively. The synthesized videos exhibit high visual fidelity and realism, demonstrating the effectiveness of ProAV-DiT in generating temporally coherent and semantically meaningful audiovisual content.
We also present results for text-to-video generation in the Fig.~\ref{text2video-landscape}. Since category names are used as text conditions during training, the model can generate corresponding videos when prompted with queries such as “a video of \textless x\textgreater”, demonstrating its capacity to generalize from text-based inputs.

% \section{The use of LLM}
% Throughout the preparation of this paper, we employed a large language model (LLM) to enhance the writing and correct grammatical mistakes.

\section{Reproducibility statement}
Implementation details, evaluation protocols, and dataset descriptions are provided in the main text and appendix. Complete proofs are also included in the main text. The full source code  will be released upon acceptance.

% \begin{table}[t]
% \centering
% \caption{MDSA Performance Comparison and Ablation Study}
% \label{tab:unified-mdsa-results-condensed-no-psnra}
% \scalebox{0.85}{
% \begin{tabular}{lcccccc}
% \toprule
% \textbf{Method/Setting} & \textbf{FVD}$\downarrow$ & \textbf{FAD}$\downarrow$ & \textbf{rFVD}$\downarrow$ & \textbf{rKVD}$\downarrow$ & \textbf{PSNR-V}$\uparrow$ \\
% \midrule
% \multicolumn{6}{c}{\cellcolor{gray!15}\textbf{Reconstruction Performance (AIST++ Dataset)}} \\
% \midrule
% MM-LDM & 53.9 & 8.9 & - & - & - \\
% MDSA-V & 30.3 & - & - & - & 31.34 \\
% MDSA-A & - & 9.4 & - & - & - \\
% \tableLineBlue \textbf{MDSA (Base)} & \textbf{18.7} & \textbf{8.7} & - & - & \textbf{32.19} \\
% \midrule
% \multicolumn{6}{c}{\cellcolor{gray!15}\textbf{Ablation Study on Generation Quality (Landscape Dataset)}} \\
% \midrule
% \textbf{MDSA (ProAV-DiT Base)} & - & 8.5 & \textbf{29.5} & \textbf{1.3} & - \\
% \midrule
% $-$ MT-selfAttn & - & 8.6 & 35.1 & 2.2 & - \\
% $-$ GCM-Attn & - & 8.7 & 39.2 & 2.3 & - \\
% $-$ Bi-Block CrossAttn & - & 8.8 & 45.6 & 2.4 & - \\
% \midrule
% $-$ finetune w/o KL loss & - & 8.9 & 60.2 & 3.2 & - \\
% $-$ w/o adversarial loss & - & 9.8 & 134.5 & 8.3 & - \\
% \bottomrule
% \end{tabular}
% }
% \end{table}

\end{document}